\documentstyle[12pt]{article}
\newcommand{\be}{\begin{equation}}
\newcommand{\en}{\end{equation}}
\newcommand{\bea}{\begin{eqnarray}}
\newcommand{\ena}{\end{eqnarray}}

\newcommand{\hbo}{\hbox to 1 true cm {\hfill } }

\newcommand{\Tr}{\hbox{Tr}}

\def\dslash{\partial\kern-.5em\slash}
\def\kslash{k\kern-.5em\slash}

\begin{document}
\vglue 1truecm

\vbox{
UNIT\"U-THEP-5$/$1993 \hfill May 4, 1993
}

\vfil
\centerline{\bf \large Non-trivial phase structure of $\phi ^{4}$-theory
   at finite temperature$^1$}

\bigskip
\centerline{ K.\ Langfeld, L.\ v.\ Smekal, H.\ Reinhardt }
\medskip
\centerline{Institut f\"ur theoretische Physik, Universit\"at
   T\"ubingen}
\centerline{D--7400 T\"ubingen, F.R.G.}
\bigskip

\vfil
\begin{abstract}

The effective potential for the local composite operator
$\phi^{2}(x)$ in  $\lambda \phi^{4}$-theory is
investigated at finite temperature in an approach based on
path-integral linearisation of the $\phi^4 $-interaction.
At zero temperature, the perturbative vacuum is unstable,
because a non-trivial phase with a scalar condensate $\langle \phi ^{2}
\rangle _{0} $ has lower effective action. Due to field renormalisation,
$\langle \lambda \phi ^{2} \rangle _{0}$ is renormalisation group
invariant and leads to the correct scale anomaly.
At a critical temperature $T_{c}$ the non-perturbative
phase becomes meta-stable implying a first order phase-transition
to the perturbative phase. The ratio
$\langle\lambda \phi^{2} \rangle _{0} / T_{c}^{2} \approx 62$
turns out to be a universal constant.

\end{abstract}

\vfil
\hrule width 5truecm
\vskip .2truecm
\begin{quote}
$^1$ Supported by DFG under contract Re $856/1 \, - \, 1$
\end{quote}
\eject
{\it 1.\ Introduction \/ }

It is of interest to study $\phi^{4}$-theory, because of its
wide field of applications~\cite{stn72}. Due to its simple
structure it also makes possible an understanding of basic features
of quantum field theory, e. g., anomalous breaking of scale
invariance~\cite{co77},  as well as being a suitable test for
non-perturbative methods, e.g.,
a variational approach~\cite{ste84} and numerical simulations
on a lattice~\cite{be83}. Lattice results
suggest that $\phi^{4}$-theory is in fact trivial (at best a free theory),
due to renormalisation effects~\cite{aiz81}.
On the other hand, many phenomenological applications~\cite{ch84} rely on a
non-trivial vacuum structure of $\phi ^{4}$-theory.
A variational approach to massive $\phi^{4}$-theory~\cite{ste84}
indicates that besides the trivial
phase, a second non-trivial one exists which escapes the triviality
bounds from lattice calculations. This phase is known as the precarious
phase since the bare coupling constant becomes (infinitesimally) negative
during renormalisation. Recently, this phase was also found for the
massless case in a modified loop expansion around the mean field~\cite{la93}.
It was shown that this expansion rapidly converges in zero dimensions
and yields renormalisation group invariant results in four
dimensions~\cite{la93}. Thereby it was observed that the so called precarious
renormalisation~\cite{ste84} is, in fact, the standard procedure and
is hidden in the usual expansion with respect to the coupling
strength~\cite{la93}, i.\ e.\ expanding the {\it precarious \/ }
renormalisation condition in powers of the renormalised coupling
constant reproduces the standard result of perturbative renormalisation.

In order to gain further insight into the vacuum structure of
$\lambda\phi^4 $-theory, it is convenient to study the effective
potential for the local composite field $\phi^2(x)$.
This is because the energy density as a function of the expectation
value of $\phi^2$ is minimised for $<\phi^2 >\not= 0 $, whereas the
effective potential of the field $\phi $ itself is
minimal at $<\phi >=0$ (also true in the non-trivial phase~\cite{la93}).

In this letter we study the temperature effects of the non-trivial phase
of $\lambda\phi^{4}$-theory. To this end we therefore calculate the effective
potential for $\phi^{2}$ in the modified loop-approach
presented in~\cite{la93} for finite temperatures. Since temperature
does not influence the renormalisation procedure, the effective
potential is renormalisation group invariant for all
temperatures, as observed in a perturbative calculation~\cite{do74}.
Due to the need of renormalisation of the operator
$\phi^2$, it turns out that
$<\lambda\phi^2 >$ is renormalisation group invariant and its
contribution to the energy density gives the correct scale anomaly
within our approximation. The vacuum energy
density of the non-trivial phase is considerably lower than the energy
density of the perturbative vacuum. The thermal expectation value of
the energy density of the non-perturbative phase
relative to the perturbative phase increases, however,
with increasing temperature, and a first order phase transition to the
perturbative phase is found
at a critical temperature $T_c$, in accordance with Landau's
theory~\cite{lan58}.
The ratio of the zero temperature condensate to the
critical temperature is found to be a universal number, as is
the ratio of the jump in the condensate at the transition point
to the critical temperature.
\vfill\eject

{\it 2.\ The effective potential of $\phi ^{2}$ \/ }

Our starting point is the Euclidean generating functional for Green's
functions, i.\ e.
\be
Z[j] \; = \; \int {\cal D} \phi \; \exp \{ -
\int_{0}^{1/T } d\tau \; \int d^{3}x \;
\bigl( \frac{1}{2} \partial _{\mu } \phi \partial _{\mu } \phi \, + \,
\frac{ m^{2} }{2} \phi ^{2} \, + \, \frac{ \lambda }{24} \phi ^{4}
\, - \, j (x) \phi ^{2} (x) \, \bigr)  \} \; ,
\label{eq:1}
\en
where $\lambda $ is the bare coupling constant and $T$ is the
temperature in units of Boltzmann's
constant. The functional integration runs over all
scalar modes satisfying periodic boundary conditions in the Euclidean
time direction. $j$ is an external source which couples linearly to
$\phi^{2}$. The effective action for $\phi^{2}$ is defined by the
Legendre transformation, i.e.,
\be
\Gamma [ \phi ^{2}_{c}] := - \ln Z[j] + \int d^{4}x \;
\phi ^{2}_{c}(x) j(x) \; , \hbo
\phi ^{2}_{c}(x) :=  \frac{ \delta \, \ln Z[j] }{ \delta j(x) } \; .
\label{eq:2}
\en
The effective potential $U(\phi ^{2}_{c})$ is obtained
by restricting $\phi _{c}$ to constant classical fields.
It is a physical quantity, e.g., its minimum value represents
the vacuum energy density~\cite{col73},
and should therefore be renormalisation group invariant.

Further access to the vacuum energy density is provided by the
scale anomaly~\cite{co77}. Conformal symmetry is broken in
the non-perturbative phase by a non-vanishing vacuum expectation value
of the trace of the energy momentum tensor, i.e.,
\be
\langle \theta _{\mu \mu } \rangle \; = \; - \frac{ \beta (
\lambda _{R} ) }{24} \; \langle \phi ^{4} \rangle \; .
\label{eq:2a}
\en
Since the vacuum is isotropic and homogeneous, we have
$\theta _{\mu \nu } = U_{0} \delta _{\mu \nu } $ with
$U_{0} = \theta _{00}$ the vacuum energy density. This  implies
\be
U_{0} \; = \; - \, \frac{1}{4}
\frac{ \beta (\lambda _{R}) }{24} \langle \phi ^{4} \rangle \; .
\label{eq:2b}
\en
For small couplings $\lambda _{R}$ the renormalisation group
$\beta $-function behaves as $\beta (\lambda _{R}) =
\beta _{0} \lambda _{R}^{2} + \ldots $ with $\beta _{0}$ a constant.
Hence if $\langle \phi ^{4} \rangle \not= 0 $, and assuming
$\langle \phi ^{4} \rangle \approx \langle \phi ^{2} \rangle ^{2}$,
leads us to regard
$\lambda _{R} \phi ^{2} $ as a physical quantity instead of $\phi
^{2}$.

Due to the $\phi^{4}$-interaction the path-integral (\ref{eq:1})
cannot be performed exactly, and one has to resort to
approximation schemes. One of these is the perturbative expansion
with respect to the coupling strength. At one-loop level, the
result for the zero-temperature effective potential for $\phi $
is~\cite{col73}
\be
U_{1-loop} \; = \; \frac{\lambda }{24} \phi _{c}^{4} \, + \,
\frac{\lambda ^{2}}{256 \pi ^{2}} \phi _{c}^{4} [ \ln
\frac{\lambda \phi _{c}^{2}}{ 2 \Lambda ^{2} } - \frac{1}{2} ] \; .
\label{eq:3}
\en
No field renormalisation is required to derive (\ref{eq:3}) implying
that $\phi _{c} $ is still renormalisation group invariant and a physical
field. It would initially seem that (\ref{eq:3}) can account
for the scale anomaly, since $U_{1-loop}(\phi _{c})$ has a minimum
for non-vanishing field. However, this minimum is an artefact of the
one-loop approximation
as already pointed out by Coleman and Weinberg, and by Coleman and
Gross~\cite{col73}. This fact is also evident from the two-loop
effective potential~\cite{ja74}, since it again has the minimum at
$\phi_{c}=0$, and significantly differs from the one-loop result for small
classical fields. For large classical fields, or large masses, the
one-loop potential scales according the standard renormalisation group
$\beta $-function which has an infra-red fixed point.

Since we would like to investigate the non-trivial phase of
$\phi ^{4}$-theory, non-per\-tur\-bative methods are required. We therefore
apply the modified loop expansion~\cite{la93}
and linearise the $\phi^{4}$-interaction by means of an auxiliary
field, i.e.,
\bea
Z[j] \; = \; \int {\cal D} \phi \; {\cal D} \chi && \! \! \! \!
\exp \{ - \int d^{4}x \; [ \, \frac{1}{2} \partial_{\mu } \phi
\partial_{\mu } \phi \, +
\label{eq:4} \\
&& \frac{6}{\lambda } \chi^{2} (x)
\, + \, [\frac{ m^{2} }{2} - i \chi (x) ] \phi ^{2}(x)
\, - \, j(x) \phi ^{2}(x) ] \; \} \; .
\nonumber
\ena
After integrating out the $\phi $-field, the modified approach~\cite{la93}
is defined by a loop expansion with respect to the field $\chi $ around
its mean field value. Even the lowest order in this expansion contains
quantum fluctuations of the $\phi $-theory, yielding reasonable results
for the effective potential of $\phi $. For the effective potential of
$\phi^{2}$, we obtain in lowest order (i.e., the mean-field approximation)
\bea
- \ln \, Z_{c} [j](T) & = &
\label{eq:5} \\
& & \int d^{4}x \; \{ - \frac{3}{2 \lambda }
(M- m ^{2} + 2j)^{2} \} \; + \; \frac{1}{2} \Tr \, \ln (
- \partial ^{2} + M ) \; ,
\nonumber
\ena
where $M$ is related to the mean field value $\chi _{0}$ by
$\chi_{0}=i (M - m^{2} + 2j ) /2 $. The mean field equation for $\chi_{0}$
can be recast into an equation for $M$, i.e.,
\be
\frac{ \delta \ln \, Z[j] }{ \delta M } \; = \; 0 \; .
\label{eq:5a}
\en
The trace in (\ref{eq:5}) extends over space-time,
and the modes satisfy periodic boundary conditions in the
Euclidean time direction. Using Schwinger's proper time regularisation
the loop contribution $L$ is
\be
L = \frac{1}{2} \Tr \, \ln (- \partial ^{2} + M ) \; = \; - \frac{V}{2}
\int \frac{ d^{3}p }{ (2\pi )^{3} } \; \sum _{n}
\int _{1 / \Lambda ^{2} } \frac{ds}{s} \; \exp \{
-s ( \vec{p}^{2} + M + ( 2\pi T )^{2} n^{2} \} \; ,
\label{eq:6}
\en
where $\Lambda $ is the proper-time cutoff and $V$ the three-space volume.
Note that the ultra-violet divergences arise from the behaviour
of the integrand in (\ref{eq:6}) at small $s$.
The integration over the three momentum is easily performed, i.e.,
\be
L  \; = \; - \frac{ V }{16 \pi ^{3/2} } \sum _{n} \int _{1 / \Lambda ^{2} }
\frac{ ds }{ s^{5/2} } \; e^{-sM} \; e^{-s
( 2\pi T )^{2} n^{2} } \; .
\label{eq:7}
\en
The divergence of the discrete sum in (\ref{eq:7}) can be extracted
by applying Poisson's formula, i.e.,
\be
\sum _{n=-\infty } ^{\infty } f(n) = \sum _{ \nu =-\infty } ^{\infty } c(\nu )
\; , \hbox to 2 true cm {\hfil with \hfil }
c(\nu ) = \int _{-\infty }^{\infty } dn \; f(n) \, e^{ i 2 \pi \nu  n } \; .
\label{eq:8}
\en
A straightforward calculation yields
\be
L \; = \; - \frac{ V_{4} }{ 32 \pi ^{2} } M^{2} \, \Gamma (-2,
\frac{M}{ \Lambda ^{2} } ) \; - \; \frac{ V_{4} }{ 32 \pi ^{2} }
\sum _{\nu \not= 0} \int _{0}^{\infty } \frac{ ds }{s ^{3} } \;
e^{-s M } \; \exp (- \frac{ 1 }{ 4s T^{2} } \nu ^{2} ) \; ,
\label{eq:9}
\en
where $\Gamma $ is the incomplete $\Gamma $-function, $V_{4}$ the
space-time volume and the limit $\Lambda \rightarrow \infty $
has been taken in the second term of the right hand side (\ref{eq:9}),
which is finite. All divergences are in the first term, which is in addition,
temperature independent.
Using the asymptotic expansion for the incomplete $\Gamma $-function
the loop contribution becomes
\be
L \; = \; \frac{ V_{4} }{32 \pi ^{2} } \{ M \Lambda ^{2} + \frac{1}{2}
M^{2} ( \ln \frac{M}{\Lambda ^{2} } - \frac{3}{2} + \gamma ) \} \; - \;
\frac{ V_{4} }{2 \pi ^{2} } \, T^{4} \,
f_{3}( \frac{ M }{ 4 T^{2} } )
\; ,
\label{eq:10}
\en
where $\gamma = 0.577...$ is the Euler-Mascheroni constant and the function
$f_{\epsilon }$ is defined by
\be
f_\epsilon (x) \; = \; \sum_{\nu \not= 0} \int _{0}^{\infty }
\frac{ ds }{ s^\epsilon } \; e^{-sx } \; e^{- \frac{  \nu ^{2} }{s} }
\label{eq:11}
\en
(where $\epsilon =3$ in (\ref{eq:10})). The term in curly brackets
in (\ref{eq:10}) is precisely the
loop contribution to the effective potential at zero temperature.
The temperature dependence is completely contained in the second term
of (\ref{eq:10}), which is finite.
As in the perturbative approach~\cite{do74}, renormalisation is unchanged,
if temperature changes.
Renormalisation is quite analogous to procedure of the zero-temperature
effective potential of $\phi $~\cite{la93}.
The key observation is that
the divergences can be absorbed into the bare parameters
$\lambda, m, j $ by setting
\bea
\frac{6}{ \lambda } \; + \; \frac{1}{16 \pi ^{2} } \,\left( \ln
\frac{\Lambda ^{2}}{\mu ^{2}} -\gamma + 1\right)  &=& \frac{6}{ \lambda _{R} }
\label{eq:12} \\
\frac{6}{\lambda } j \, - \, \frac{3 m^{2} }{\lambda } \; - \;
\frac{1}{32 \pi ^{2} } \Lambda ^{2} &=& \frac{6}{ \lambda _{R} } j_{R}
\, - \, \frac{ 3 m_{R}^{2} }{ \lambda _{R} }
\label{eq:13} \\
j \; - \; m^{2} &=& 0 \; ,
\label{eq:14}
\ena
where $\mu $ is an arbitrary renormalisation  scale. Later,
we will check that physical quantities do not depend on $\mu $.
Note that the mean field variable $M$ has been introduced in such a way
that renormalisation is possible without $M$ picking up divergences.
In fact, as will be shown later, $M$ is a renormalisation group invariant.
In the following we confine ourselves to the
massless case $(m_{R}=0)$. The bare
parameters are uniquely defined, in (\ref{eq:12})-(\ref{eq:14}),
as functions of the renormalised ones and
the regulator. The renormalisation of the coupling
constant (\ref{eq:12}) is essentially the same as in \cite{la93}, implying a
universal renormalisation procedure in the calculation of the
effective potential of $\phi $ and $\phi^{2}$. Besides coupling constant
renormalisation, mass {\it and} field renormalisation (\ref{eq:13},
\ref{eq:14}) is also necessary to absorb all divergencies in (\ref{eq:5}).
This is different to the case of the effective potential of $\phi $
on its own, where the field does not require renormalisation at this
level. The situation here is analogous to the Gaussian effective potential
approach to $\lambda \phi^4 $-theory (see refs. \cite{ste84,soto87}),
where besides the "precarious phase", a second non-trivial phase
(called "autonomous phase" in~\cite{ste84}) would
seem to exist, allowing for an infinite shift of the field variable.
It was shown, however, that field renormalisation does not truly
give a new phase, but rather what is the effective potential of the
composite field $\lambda\phi^2$~\cite{soto87}.

Introducing the renormalised quantities
in (\ref{eq:10}), the generating functional (\ref{eq:5}) becomes
\bea
-\frac{1}{ V_{4} } \ln \, Z[j](T) &=&
- \frac{3}{2 \lambda _{R} } M^{2} \, - \,
\frac{6}{\lambda _{R} } M j_{R} \, + \, \frac{ \alpha }{ 2 }
M^{2} ( \ln \frac{M}{\mu ^{2} } - \frac{1}{2} )
\label{eq:15} \\
&-&  16 \alpha \, T^{4} \,
f_{3}( \frac{ M }{ 4 T^{2} } ) \; ,
\nonumber
\ena
where $\alpha = 1/ 32 \pi ^{2}$. Here $M$ is defined
by the mean field equation (\ref{eq:5a}), which in terms of the
renormalised quantities reads
\be
- \frac{3}{\lambda _{R} } M \, - \, \frac{6}{\lambda _{R} } j_{R}
\, + \, \alpha M  \ln \frac{M}{\mu ^{2} } \, + \,
4 \alpha \, T^{2} \,
f_{2}( \frac{ M }{ 4 T^{2} } ) \; = \; 0
\label{eq:16}
\en
where $f_{2}$ is defined by (\ref{eq:11}).
In order to perform the Legendre transformation (\ref{eq:2}) we first
obtain the classical field, i.e.,
\be
\phi ^{2}_{c} \; = \; \frac{1}{ V_{4} }
\frac{ \partial \ln Z[j_{R}] }{\partial j_{R} } \; = \;
\frac{6}{\lambda _{R}} M \; .
\label{eq:17}
\en
The effective potential (\ref{eq:2}), is therefore given by
\be
U(\phi _{c}^{2}) \; = \; \frac{\alpha }{2} M^{2} (
\ln \frac{M}{\mu ^{2} } - \frac{1}{2} ) \; - \;
\frac{3}{2 \lambda _{R} } M ^{2} \; - \; 16 \alpha T^{4} \,
f_{3}( \frac{ M }{ 4 T^{2} } ) \,
\vert _{ M = \frac{\lambda _{R} }{6} \phi _{c}^{2} } \; .
\label{eq:18}
\en
We have obtained a finite and renormalisation group invariant result,
since a change in the scale $\mu $ can be absorbed by
a redefinition of the renormalised coupling constant $\lambda _{R}$ according
(\ref{eq:12}). The renormalisation group $\beta $-function
\be
\beta (\lambda _{R}) \; = \; \beta _{0} \, \lambda _{R}^{2} \; ,
\hbox to 2 true cm {\hfil with \hfil }
\beta _{0} := \frac{2 \alpha }{3}
\label{eq:19}
\en
exhibits an infrared stable fixed point as indicated by a
perturbative calculation~\cite{col73}. Note that since our
approach is intrinsically non-perturbative, (\ref{eq:19})
need not agree with the $\beta $-function from the perturbative
calculation. Furthermore, we see that $M= \lambda _{R}
\phi _{c}^{2}/6 $ rather than $\phi _{c}^{2}$ is renormalisation group
invariant, as suggested by the scale anomaly. Due to the
need of field renormalisation $\phi _{c}^{2} $ itself has lost
its physical meaning.

Figure 1 shows the effective potential (\ref{eq:18}) as
a function of $\lambda _{R} \phi _{c}^{2}$ for various temperatures.
At zero temperature it develops a minimum $U_{0}$ at a non-vanishing
value of the condensate $M_{0}$, i.e.,
\be
M_{0} \; = \; \mu ^{2}
\exp \{ \frac{2}{\beta _{0} \lambda _{R} } \} \;,
\hbo U_{0} \; = \; - \frac{\alpha }{4}
\mu ^{4}  \exp \{ \frac{4}{\beta _{0} \lambda _{R} } \} \; .
\label{eq:20}
\en
This minimum corresponds to a {\it non-trivial} vacuum -
at zero temperature it has lower energy density
than the perturbative state $(\phi_{c}^{2}=0)$.
The vacuum energy density of the non-perturbative phase is
\be
U_{0} \; = \; -{\alpha \over 4}M_0^2 \; =\; - \frac{\alpha }{144}
\lambda^{2}_{R} (\phi_{c}^{2})^{2}
\; \rightarrow \; - \frac{1}{4}
\frac{ \beta (\lambda _{R}) }{24} \langle :\!\phi ^{4}\!: \rangle \; .
\label{eq:21}
\en
This is precisely the result obtained in \cite{la93} for
$U(\phi_c=0)$ in the non-trivial phase, furthermore it gives the
correct scale anomaly (\ref{eq:2b}) obtained by general arguments.
Note that in the effective $\chi $-theory, obtained from (\ref{eq:4})
by integrating out $\phi $, the trace anomaly already shows up at
tree level. It is therefore not so surprising that the trace anomaly
survives in the mean-field approximation.

Increasing the temperature shifts the scalar condensate to smaller values
and lowers the difference of the energy density of perturbative and
non-perturbative phase. At a critical temperature $T_{c}$,
the non-trivial phase becomes degenerate with the perturbative
one (at $\phi _{c}^{2}=0$). If the temperature is increased further,
the non-trivial phase becomes meta-stable and a first order
phase transition to the trivial phase at $\phi _{c}^{2}=0$ can
occur either by quantum or statistical fluctuations. This phase
transition implies that the scalar condensate
as a function of temperature has a discontinuity at the critical
temperature.

\bigskip
{\it 3.\ At the phase-transition \/ }

We first investigate the change of the scalar condensate with
temperature. The condensate $M _{v} = \lambda _{R} \phi ^{2}_{c} /6
\vert _{min}$ is obtained by solving the equation of motion (\ref{eq:16})
for zero external source $(j_{R}=0)$ which is
\be
- \frac{3}{\lambda _{R} }  \, + \, \alpha  \ln \frac{ M_{v} }{\mu ^{2} }
\, + \, \frac{ \alpha }{ x  } f_{2}( x ) \; = \; 0 \; ,
\label{eq:22}
\en
where $x:= M_{v} /4T^{2} $. In order to remove the renormalisation
point dependence we subtract the corresponding equation for the
zero temperature condensate $M_{0}$ given by
\be
- \frac{3}{\lambda _{R} }  \, + \, \alpha   \ln \frac{ M_{0} }{\mu^{2}}
 \; = \; 0
\label{eq:23}
\en
yielding
\be
\ln \frac{ M_{v} }{ M_{0} } \; + \; \frac{1}{x} f_{2}(x) \; = \; 0 \; .
\label{eq:24}
\en
This equation allows us
to calculate $M_{v}$, in terms of its zero temperature value,
as function of temperature. A numerical solution of (\ref{eq:24}) is
displayed in figure 2.

It is instructive to relate the jump in the condensate and the critical
temperature $T_{c}$ to a fundamental observable like the
scalar condensate at zero temperature $M_{0}=\langle \lambda _{R}
\phi ^{2} /6 \rangle $. At the critical temperature, the energy
density of the perturbative thermal ground state $U(\phi ^{2}_{c}=0)$ is equal
to that of the non-trivial phase $U(\phi ^{2}_{c \; min } )$.

The effective potential at zero field is obtained from (\ref{eq:18})
by a direct calculation, i.\ e.
\be
U_{per} \; := \; U(\phi _{c}^{2}=0) \; = \; - 16 \alpha T^{4}
f_{3}(0) \; = \; - 32 \alpha T^{4} \zeta (4) \; ,
\label{eq:25}
\en
where $\zeta $ is the Riemann $\zeta $-function.
Equating the energy density of the perturbative thermal state
$U_{per}$ to the one  of the non-trivial phase (with non-zero
condensate) we obtain
\be
- \frac{3}{2 \lambda _{R} } \; + \; \frac{\alpha }{2} ( \ln
\frac{M_{v} }{\mu ^{2} } - \frac{1}{2} ) \; - \;
\frac{ \alpha }{ x^{2} } f_{3}(x) \; = \; - \frac{ 2 \alpha }{x^{2}}
\zeta (4) \; .
\label{eq:26}
\en
In order to eliminate the renormalisation point dependence we
replace $\mu $ by $M_{0}$ by using (\ref{eq:20}),
yielding
%
%
%
\be
\ln \frac{ M_{v} }{ M_{0} } \; = \; \frac{1}{2} \; + \; \frac{ 2 }{ x^{2} }
[ f_{3}(x) - 2 \zeta (4) ] \; .
\label{eq:28}
\en
Combining this equation with (\ref{eq:24}) we finally obtain
\be
\frac{1}{2} \; + \; \frac{2}{x^{2}} [ f_{3}(x) - 2 \zeta (4) ]
\; + \; \frac{1}{x} f_{2}(x) \; = \; 0 \; .
\label{eq:29}
\en
This relation allows us to calculate the ratio of the jump of condensate
to the square of the critical temperature. A numerical solution
of (\ref{eq:29}) yields
\be
\frac{ M_{v}( T_{c} ) }{  T_{c} ^{2} } \; = \; 4 x \; = \; 9.30472... \; .
\label{eq:30}
\en
Furthermore the critical temperature is related to the zero temperature
condensate by
\be
\ln \frac{ M_{0} }{ 4 T_{c} ^{2} } \; = \;
\ln x \; - \; \frac{1}{2} \; - \; \frac{2}{x^{2}} [ f_{3}(x) - 2 \zeta (4)]
\label{eq:31}
\en
implying
\be
\frac{ M_{0} }{ T_{c} ^{2} } \; = \; 10.29134... \; .
\label{eq:32}
\en
A scalar condensate $M_{0} \approx (100 \, \hbox{MeV} \, )^{2}$
at zero temperature would imply a transition temperature of
$T_{c} \approx 31 \, $MeV.

We have shown that at a critical temperature the mean-field effective
potential $\Gamma [ \phi ^{2} ]$ exhibits a first order phase
transition from a non-trivial phase realised at zero temperature
to a trivial (perturbative) phase realised at high temperatures.
Cosmological consequences of this phase transition will be the
subject of further investigations.

\bigskip
\leftline{\bf Acknowledgements: }

We want to thank R.\ F.\ Langbein for carefully reading this
manuscript and helpful remarks.
\medskip

\vspace{2 true cm}
\leftline{\bf Figure captions }

\bigskip
{\bf figure 1 } The effective potential as function of the scalar condensate
   at various temperatures. A large coupling strength $\lambda _{R}(\mu)
   =1000$ was chosen for aesthetic reasons.

\bigskip
{\bf figure 2 } The scalar condensate as function of temperature.

\end{document}